\documentclass[usenatbib,referee]{mn2e}
\usepackage{amsmath}
\usepackage{graphicx}
\usepackage{epsfig}
\usepackage{txfonts}
\usepackage{color}
\usepackage{natbib}

\title[The outcomes of CO WDs accreting CO-rich material]
{The outcomes of carbon-oxygen white dwarfs accreting CO-rich material}
\author[C. Wu, B. Wang and D. Liu]
{Chengyuan Wu$^{\rm 1,2,3}$\thanks{E-mail:wcy@ynao.ac.cn}, Bo Wang$^{\rm 1,2,3}$\thanks{E-mail:wangbo@ynao.ac.cn} and Dongdong Liu$^{\rm 1,2,3}$\\
$^1$Key Laboratory for the Structure and Evolution of Celestial Objects, Yunnan Observatories, CAS, Kunming 650216, China\\
$^2$University of Chinese Academy of Sciences, Beijing 100049, China\\
$^3$Centre for Astronomical Mega-Science, CAS, Beijing 100012, China}

\begin{document}
%\date{Accepted. Received}
\date{}
\pagerange{\pageref{firstpage}--\pageref{lastpage}} \pubyear{2018}
\maketitle

\label{firstpage}

\begin{abstract}\label{0. abstract}
The double-degenerate model, involving the merger of double carbon-oxygen white dwarfs (CO WDs), is one of the two classic models for the progenitors of type Ia supernovae (SNe Ia). Previous studies suggested that off-centre carbon burning would occur if the mass-accretion rate ($\dot{M}_{\rm acc}$) is relatively high during the merging process, leading to the formation of oxygen-neon (ONe) cores that may collapse into neutron stars. However, the off-centre carbon burning is still incompletely understood, especially when the inwardly propagating burning wave reaches the centre. In this paper, we aim to investigate the propagating characteristics of burning waves and the subsequently evolutionary outcomes of these CO cores. We simulated the long-term evolution of CO WDs that accrete CO-rich material by employing the stellar evolution code MESA on the basis of the thick-disc assumption. We found that the final outcomes of CO WDs strongly depend on $\dot{M}_{\rm acc}$ (${M}_\odot\,\mbox{yr}^{-1}$) based on the thick-disc assumption, which can be divided into four regions: (1) explosive carbon ignition in the centre, then SNe Ia ($\dot{M}_{\rm acc}<2.45\times10^{-6}$); (2) OSi cores, then neutron stars ($2.45\times10^{-6}\lesssim\dot{M}_{\rm acc}<4.5\times10^{-6}$); (3) ONe cores, then e-capture SNe ($4.5\times10^{-6}\lesssim\dot{M}_{\rm acc}<1.05\times10^{-5}$); (4) off-centre oxygen and neon ignition, then off-centre explosion or Si-Fe cores ($\dot{M}_{\rm acc}\gtrsim1.05\times10^{-5}$). Our results indicate that the final fates of double CO WD mergers are strongly dependent on the merging processes (e.g. slow merger, fast merger, composite merger, violent merger, etc.).
\end{abstract}

\begin{keywords}
binaries: close -- stars: evolution -- supernovae: general -- white dwarfs
\end{keywords}

\section{Introduction} \label{1. Introduction}

Type Ia supernovae (SNe Ia) are used as cosmological distance indicators successfully since they have uniformly high luminosities (e.g. Riess et al. 1998; Perlmutter et al. 1999). They originate from thermonuclear explosions of carbon-oxygen white dwarfs (CO WDs; e.g. Hoyle \& Fowler 1960; Woosley, Taam \& Weaver 1986). However, their progenitor systems are still not well determined (e.g. Wang \& Han 2012; Wang et al. 2013; Maoz, Mannucci \& Nelemans 2014). At present, two progenitor models of SNe Ia have been widely discussed, i.e. the single-degenerate model and the double-degenerate model (e.g. Whelan \& Iben 1973; Nomoto, Thielemann \& Yokoi 1984; Webbink 1984; Iben \& Tutukov 1984). In the single-degenerate model, a CO WD accretes material from its non-degenerate companion (e.g. a main-sequence star, a red-giant star or a helium star), and finally a SN Ia may be produced when the WD grows in mass up to the Chandrasekhar-mass limit (${M}_{\rm Ch}$; e.g. Hachisu, Kato \& Nomoto 1996; Li \& van den Heuvel 1997; Langer et al. 2000; Han \& Podsiadlowski 2004, 2006; Meng, Chen \& Han 2009; Wang et al. 2009). In the double-degenerate model, two CO WDs merge via the orbital angular momentum loss caused by gravitational wave radiation. It has been suggested that SN Ia can be produced if the total mass of the double WDs exceed ${M}_{\rm Ch}$ (e.g. Nelemans et al. 2001; Ruiter, Belczynski \& Fryer 2009; Soker 2018).

The classic double-degenerate model is supported by some observational evidence (e.g. Livio \& Mazzali 2018; Wang 2018). For example, no hydrogen and helium lines in nebular spectra of most SNe Ia (e.g. Leonard 2007; Ganeshalingam, Li \& Filippenko 2011), no detections of ejecta-companion interaction in some SNe Ia (e.g. Olling et al. 2015), no confirmed observational evidence of surviving companion stars of SNe Ia (e.g. Badenes et al. 2007; Edwards, Pagnotta \& Schaefer 2012; Graham et al. 2015), no signature of early radio emission (e.g. Hancock, Gaensler \& Murphy 2011; Horesh et al. 2012), the observed evidence for the existence of some superluminous events (e.g. Howell et al. 2006; Hichen et al. 2007; Scalzo et al. 2010), etc. In addition, the double-degenerate model can reproduce the observed birthrates and delay-time distributions of SNe Ia (e.g. Toonen, Nelemans \& Portegies Zwart 2012; Ruiter et al. 2013; Liu et al. 2016; Yungelson \& Kuranov 2017; Liu, Wang \& Han 2018).

The mergers of double WDs have been widely investigated both in stellar evolution studies of the outcomes (e.g. Nomoto \& Iben 1985; Saio \& Nomoto 1985, 1998; Piersanti et al. 2003a,b; Yoon, Podsiadlowski \& Rosswog 2007) and in hydrodynamical studies of the merger process itself (e.g. Pakmor et al. 2010; Dan et al. 2012, 2014). Previous works investigated the merging process of double CO WDs based on the thick-disc assumption (slow merger), in which the less massive WD could be destroyed and then forms a pressure-supported disc around the massive WD (the mass-accretion rate from the disc to the massive WD is assumed to be constant and the accreting process can last for millions of years; e.g. Nomoto \& Iben 1985; Kawai \& Saio 1987; Saio \& Nomoto 1985, 1998, 2004; Saio \& Jeffery 2000, 2002). Saio \& Nomoto (1985) suggested that the off-centre carbon ignition would occur if the mass-accretion rate is higher than a critical rate ($2-4\times10^{-6}\,{M}_\odot\,\mbox{yr}^{-1}$; see also Nomoto \& Iben 1985; Kawai \& Saio 1987; Martin, Tout \& Lesaffre 2006). Subsequently, the inwardly propagating carbon flame would reach the centre, transforming the CO WD into an ONe core, resulting in the formation of a neutron star rather than a SN Ia (e.g. Timmes, Woosley \& Taam 1994; Podsiadlowski et al. 2004; Shen et al. 2012; Schwab, Quataert \& Bildsten 2015; Brooks et al. 2016; Wang, Podsiadlowski \& Han 2017). Recently, Schwab, Quataert \& Kasen (2016) studied the remnants of double CO WD mergers (fast merger), and found that the evolution of super-${M}_{\rm Ch}$ remnants are somewhat like super-AGB stars. They suggested that off-centre neon ignitions would occur after the inwardly propagating carbon flames reach their centres (see also Jones et al. 2013; Jones, Hirschi \& Nomoto 2014).

Although the rapid mass-accretion process is more likely to trigger the off-centre carbon ignition, some studies suggested that the mergers of double CO WDs can produce SNe Ia in some parameter ranges (e.g. Piersanti et al. 2003a,b; Pakmor et al. 2010; Sato et al. 2015). Yoon, Podsiadlowski \& Rosswog (2007) investigated the remnant evolution of double CO WD merger based on the composite merger assumption (composite merger), in which the merging system includes a slowly rotating cold core, a rapidly rotating hot corona and a centrifugally supported disc. They found that SNe Ia can be produced if the post-merger evolution satisfies two criteria: (1) the spin-down timescale of the central event needs to be longer than that of neutrino loss; (2) the mass-accretion rate from the centrifugally supported disc needs to be relatively low. Moreover, the violent WD merger scenario may also produce SNe Ia, in which two WDs merge violently if the mass ratio is large enough (e.g. $q>0.8$), leading to the occurrence of prompt detonation during the merger process (e.g. Pakmor et al. 2010, 2011, 2012).

In the previous simulations, the inwardly propagating carbon flame has not reached the centre (e.g. Saio \& Nomoto 1985, 1998, 2004), which may lead to some uncertainties in studying the final outcomes of double CO WD mergers. In this work, we aim to study the inwardly propagating carbon flame approaching the centre and the subsequent evolution. We investigate the propagating burning waves and give the final outcomes of CO WD mergers based on the assumption of the thick-disc model. In Sect. 2, we provide our basic assumptions and methods of numerical simulations. The results and examples of our simulations are shown in Sect. 3. Finally, we present discussion in Sect. 4 and summary in Sect. 5.

\section{Numerical Methods}\label{Methods}

We use \texttt{Modules for Experiments in Stellar Astrophysics} (MESA, version 7624; see Paxton et al. 2011, 2013, 2015) to simulate the long-term evolution of CO WDs accreting CO-rich material. The default OPAL opacity is adopted (e.g. Iglesias \& Rogers 1993, 1996), and the nuclear reaction network used in our simulations is \texttt{co\_burn.net}, which mainly includes the isotopes needed for carbon, oxygen, neon and magnesium burning (e.g. $^{\rm 12}{\rm C}$, $^{\rm 16}{\rm O}$, $^{\rm 20}{\rm Ne}$, $^{\rm 24}{\rm Mg}$, $^{\rm 28}{\rm Si}$). This nuclear reaction network is coupled by more than $50$ reactions, mainly including reactions as follows:

\begin{equation}
^{\rm 12}{\rm C} + ^{\rm 12}{\rm C}\,\rightarrow\,^{\rm 20}{\rm Ne} + ^{\rm 4}{\rm He},
\end{equation}
\begin{equation}
^{\rm 12}{\rm C} + ^{\rm 16}{\rm O}\,\rightarrow\,^{\rm 24}{\rm Mg} + ^{\rm 4}{\rm He},
\end{equation}
\begin{equation}
^{\rm 12}{\rm C} + ^{\rm 16}{\rm O}\,\rightarrow\,^{\rm 28}{\rm Si},
\end{equation}
\begin{equation}
^{\rm 20}{\rm Ne} + \gamma\,\rightarrow\,^{\rm 16}{\rm O} + ^{\rm 4}{\rm He},
\end{equation}
\begin{equation}
^{\rm 20}{\rm Ne} + ^{\rm 4}{\rm He}\,\rightarrow\,^{\rm 24}{\rm Mg},
\end{equation}
and
\begin{equation}
^{\rm 24}{\rm Mg} + ^{\rm 4}{\rm He}\,\rightarrow\,^{\rm 28}{\rm Si}.
\end{equation}
This nuclear network includes all the main isotopes and nuclear reactions for off-centre carbon burning, which can effectively save the calculation time. However, this nuclear network does not include the iron-group elements and might not be suitable for reactions in extremely high temperature. We will discuss the influence of different nuclear networks in Sect. 4.

Firstly, we construct a series of hot CO cores ranging from $0.5-1.0{M}_\odot$. In our simulations, all the CO cores are built in idealized methods, in which the mass fractions of $^{\rm 12}{\rm C}$ and $^{\rm 16}{\rm O}$ are both $50\%$. During the formation process of hot cores, all the nuclear reactions are neglected, resulting in that the cores have uniform elemental abundance distribution from the inside out. The cores are cooled down until the cooling time is in the range of $10^{5}-10^{7}\,\mbox{yr}$ after the cores enter the cooling phase. At this moment, the initial WD models have been constructed, in which the effective temperatures range from $7.4\times10^{4}$ to $3.5\times10^{5}\,{\rm K}$. Secondly, we use the WD models to accrete CO material with the same elemental abundance by adopting constant mass-accretion rates based on the thick-disc assumption (e.g. Saio \& Nomoto 1985; Nomoto \& Iben 1985). The mass-accretion rates ($\dot{M}_{\rm acc}$) in our simulations are in the range of $10^{-6}-2\times10^{-5}\,{M}_\odot\,\mbox{yr}^{-1}$. In the present work, we did not consider the convective overshoot mixing, but some recent studies indicate that this process can influence the propagation of burning waves (e.g. Denissenkov et al. 2013a,b).

\section{Numerical Results}\label{Results}

\begin{figure}
\begin{center}
\epsfig{file=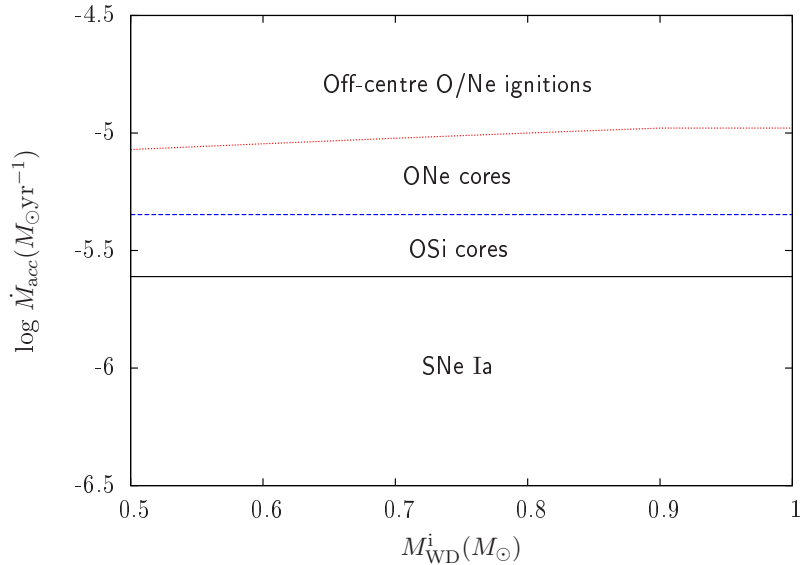,angle=0,width=12.2cm}
 \caption{The final evolutionary outcomes of different CO WDs in a wide range of mass-accretion rates. Three lines present boundaries of different regions.}
  \end{center}
\end{figure}

We evolved a series of CO WDs by accreting CO-rich material, and found that the final outcomes of the accreting WDs are sensitive to the accretion rates. The relationship between the final outcomes and $\dot{M}_{\rm acc}$ for different initial WD masses (${M}_{\rm WD}^{\rm i}$) are shown in Fig.\,1. Four regions in this figure represent different evolutionary outcomes of CO WDs:

(1) Explosive carbon ignition in the centre, then SNe Ia ($\dot{M}_{\rm acc}<2.45\times10^{-6}\,{M}_\odot\,\mbox{yr}^{-1}$). In this case, the accreting WDs will increase their masses to $M_{\rm Ch}$, and eventually the explosive carbon ignitions would occur in the centres of WDs, leading to the formation of SNe Ia.

(2) OSi cores ($2.45\times10^{-6}\,{M}_\odot\,\mbox{yr}^{-1}\lesssim\dot{M}_{\rm acc}<4.5\times10^{-6}\,{M}_\odot\,\mbox{yr}^{-1}$). In this case, the off-centre burning waves transform carbon into silicon directly, resulting in the formation of OSi cores.

(3) ONe cores ($4.5\times10^{-6}\,{M}_\odot\,\mbox{yr}^{-1}\lesssim\dot{M}_{\rm acc}<1.05\times10^{-5}\,{M}_\odot\,\mbox{yr}^{-1}$). The evolutionary characteristics of accreting WDs in this case are similar to those in Saio \& Nomoto (1985); the inwardly propagating burning waves transform carbon into neon, leading to the formation of ONe cores.

(4) Off-centre O/Ne ignitions ($\dot{M}_{\rm acc}\gtrsim1.05\times10^{-5}$). Firstly, the ONe cores are formed via off-centre carbon burning. Secondly, off-centre neon is ignited and then produce the inwardly propagating neon flames. Finally, the dynamical processes are triggered when the neon flames move close to their centres.

Our present work implies that the initial WD masses have almost no influence on the critical lines, which means that the final outcomes for various WDs are primarily related to the mass-accretion rates. In Table\,1, we summarize the outcomes of CO WDs accreting CO-rich material. The representative examples of these cases are given as follows.

\begin{table*}
\centering
\caption{Outcomes of CO WDs accreting CO-rich material. Notes: ${M}_{\rm WD}^{\rm i}$ = initial WD mass; $\log\dot{M}_{\rm acc}$ = mass-accretion rate in logarithmic form; ${M}_{\rm WD}^{\rm f}$ = the final core mass; Outcomes = possible final outcomes of the core. All the initial CO WD models in our simulations have the same cooling time $t_{\rm cool}=1.0\times10^{6}\,\mbox{yr}$. ${M}_{\rm WD}^{\rm f}$ is given under the assumption that mass-accretion process is stopped if burning wave has reached the centre or explosion has occurred.}
\begin{tabular}{|c|c|c|c|}     % 8 columns
\hline
 ${M}_{\rm WD}^{\rm i} ({M}_\odot)$ & $\log\dot{M}_{\rm acc} ({M}_\odot\,\mbox{yr}^{-1})$  &${M}_{\rm WD}^{\rm f} ({M}_\odot)$  &Outcomes\\
 \hline
 0.5                      &$2.40\times10^{-6}$  &$1.3738$  &SN Ia\\
                          &$2.45\times10^{-6}$  &$1.3719$  &OSi core $\rightarrow$ OSi WD/CCSN\\
                          &$4.00\times10^{-6}$  &--        &OSi core $\rightarrow$ OSi WD/CCSN\\
                          &$4.50\times10^{-6}$  &--        &OSi core $\rightarrow$ OSi WD/CCSN\\
                          &$8.00\times10^{-6}$  &$1.3644$  &ONe core $\rightarrow$ ECSN\\
                          &$8.50\times10^{-6}$  &$1.3796$  &ONe core $\rightarrow$ off-centre neon ignition $\rightarrow$ dynamical process\\
 \hline
 0.6                       &$2.40\times10^{-6}$  &$1.3746$  &SN Ia\\
                           &$2.45\times10^{-6}$  &$1.3713$  &OSi core $\rightarrow$ OSi WD/CCSN\\
                           &$4.00\times10^{-6}$  &--        &OSi core $\rightarrow$ OSi WD/CCSN\\
                           &$4.50\times10^{-6}$  &--        &OSi core $\rightarrow$ OSi WD/CCSN\\
                           &$8.50\times10^{-6}$  &$1.3612$  &ONe core $\rightarrow$ ECSN\\
                           &$9.00\times10^{-6}$  &$1.3786$  &ONe core $\rightarrow$ off-centre neon ignition $\rightarrow$ dynamical process\\
 \hline
 0.7                       &$2.40\times10^{-6}$  &$1.3727$  &SN Ia\\
                           &$2.45\times10^{-6}$  &$1.3714$  &OSi core $\rightarrow$ OSi WD/CCSN\\
                           &$4.00\times10^{-6}$  &--        &OSi core $\rightarrow$ OSi WD/CCSN\\
                           &$4.50\times10^{-6}$  &--        &OSi core $\rightarrow$ OSi WD/CCSN\\
                           &$9.00\times10^{-6}$  &$1.3588$  &ONe core $\rightarrow$ ECSN\\
                           &$9.50\times10^{-6}$  &$1.3728$  &ONe core $\rightarrow$ off-centre neon ignition $\rightarrow$ dynamical process\\
                           &$1.00\times10^{-5}$  &$1.3854$  &ONe core $\rightarrow$ off-centre neon ignition $\rightarrow$ dynamical process\\
                           &$1.05\times10^{-5}$  &$1.3984$  &ONe core $\rightarrow$ off-centre neon ignition $\rightarrow$ dynamical process\\
                           &$1.90\times10^{-5}$  &$1.6007$  &ONe core $\rightarrow$ off-centre neon ignition $\rightarrow$ dynamical process\\
 \hline
 0.8                       &$2.40\times10^{-6}$  &$1.3744$  &SN Ia\\
                           &$2.45\times10^{-6}$  &$1.3715$  &OSi core $\rightarrow$ OSi WD/CCSN\\
                           &$4.00\times10^{-6}$  &--        &OSi core $\rightarrow$ OSi WD/CCSN\\
                           &$4.50\times10^{-6}$  &--        &OSi core $\rightarrow$ OSi WD/CCSN\\
                           &$9.50\times10^{-6}$  &$1.3591$  &ONe core $\rightarrow$ ECSN\\
                           &$1.00\times10^{-5}$  &$1.3707$  &ONe core $\rightarrow$ off-centre neon ignition $\rightarrow$ dynamical process\\
 \hline
 0.9                       &$2.40\times10^{-6}$  &$1.3747$  &SN Ia\\
                           &$2.45\times10^{-6}$  &$1.3719$  &OSi core $\rightarrow$ OSi WD/CCSN\\
                           &$4.00\times10^{-6}$  &$1.2875$  &OSi core $\rightarrow$ OSi WD/CCSN\\
                           &$4.50\times10^{-6}$  &--        &OSi core $\rightarrow$ OSi WD/CCSN\\
                           &$7.00\times10^{-6}$  &$1.3175$  &ONe core $\rightarrow$ ECSN\\
                           &$1.00\times10^{-5}$  &$1.3680$  &ONe core $\rightarrow$ ECSN\\
                           &$1.05\times10^{-5}$  &$1.3782$  &ONe core $\rightarrow$ off-centre neon ignition $\rightarrow$ dynamical process\\
                           &$1.10\times10^{-5}$  &$1.3878$  &ONe core $\rightarrow$ off-centre neon ignition $\rightarrow$ dynamical process\\
                           &$1.20\times10^{-5}$  &$1.4063$  &ONe core $\rightarrow$ off-centre neon ignition $\rightarrow$ dynamical process\\
                           &$1.50\times10^{-5}$  &$1.4640$  &ONe core $\rightarrow$ off-centre neon ignition $\rightarrow$ dynamical process\\
                           &$1.80\times10^{-5}$  &$1.5201$  &ONe core $\rightarrow$ off-centre neon ignition $\rightarrow$ dynamical process\\
                           &$1.90\times10^{-5}$  &$1.5380$  &ONe core $\rightarrow$ off-centre neon ignition $\rightarrow$ dynamical process\\
 \hline
 1.0                       &$2.40\times10^{-6}$  &$1.3748$  &SN Ia\\
                           &$2.45\times10^{-6}$  &$1.3712$  &OSi core $\rightarrow$ OSi WD/CCSN\\
                           &$4.00\times10^{-6}$  &--        &OSi core $\rightarrow$ OSi WD/CCSN\\
                           &$4.50\times10^{-6}$  &--        &OSi core $\rightarrow$ OSi WD/CCSN\\
                           &$1.00\times10^{-5}$  &$1.3561$  &ONe core $\rightarrow$ ECSN\\
                           &$1.05\times10^{-5}$  &$1.3669$  &ONe core $\rightarrow$ off-centre neon ignition $\rightarrow$ dynamical process\\
                           &$1.90\times10^{-5}$  &$1.5267$  &ONe core $\rightarrow$ off-centre neon ignition $\rightarrow$ dynamical process\\
 \hline
\end{tabular}
\end{table*}

\subsection{SNe Ia}

\begin{figure}
\begin{center}
\epsfig{file=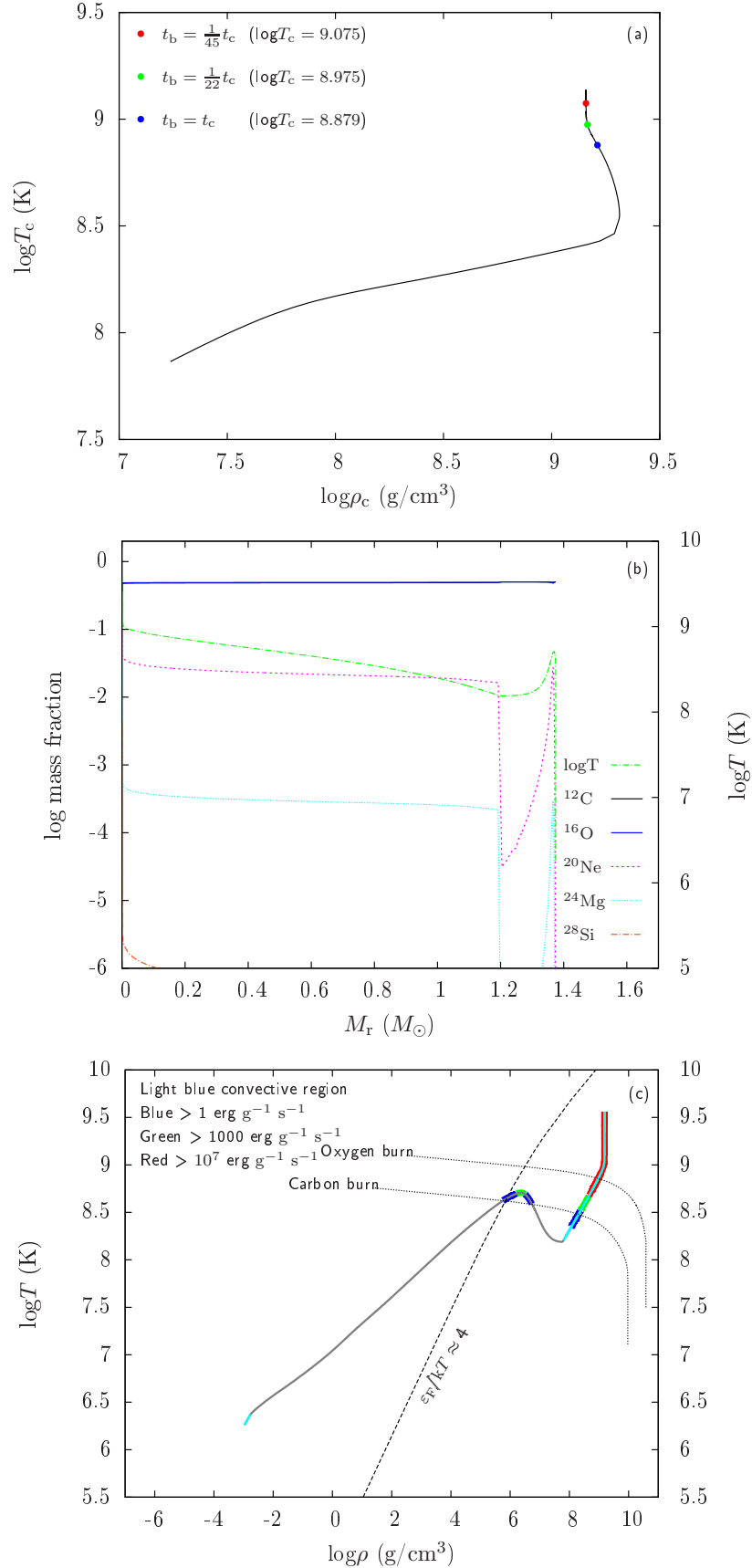,angle=0,width=11.9cm}
 \caption{An example of producing SNe Ia, in which ${M}_{\rm WD}^{\rm i}=0.9{M}_\odot$ and $\dot{M}_{\rm acc}=2.4\times10^{-6}\,{M}_\odot\,\mbox{yr}^{-1}$. Panel (a): the central density-temperature profile during the evolution, in which three filled circles represent different points where $t_{\rm b}=1,\frac{1}{22},\frac{1}{45}t_{\rm c}$. Panel (b): the elemental abundance distribution at the moment of explosive carbon ignition. Panel (c): the density-temperature profile at the moment of explosive carbon ignition.}
  \end{center}
\end{figure}

If $\dot{M}_{\rm acc}$ is lower than $2.45\times10^{-6}\,{M}_\odot\,\mbox{yr}^{-1}$, the accreting WDs will grow in mass up to ${M}_{\rm Ch}$ and trigger the explosive carbon ignitions in their centres. Fig.\,2 shows a representative example of producing SNe Ia, in which ${M}_{\rm WD}^{\rm i}=0.9{M}_\odot$ and $\dot{M}_{\rm acc}=2.4\times10^{-6}\,{M}_\odot\,\mbox{yr}^{-1}$ (in the region of ``SNe Ia''). At the beginning of evolution, carbon in the shell is heated up owing to the mass-accretion process. The CO WD can increase its mass up to ${M}_{\rm Ch}$ before the occurrence of off-centre carbon ignition because of the relatively low mass-accretion rate.

It has been suggested that SN Ia explosion will occur when the mass of CO WD is close to ${M}_{\rm Ch}$, but the critical point of explosive carbon ignition is still under debate (e.g. Woosley, Wunsch \& Kuhlen 2004; Lesaffre et al. 2006; Chen, Han \& Meng 2014; Wang, Podsiadlowski \& Han 2017). Wunsch \& Woosley (2004) suggested that the explosive carbon ignition occurs when the nuclear burning timescale ($t_{\rm b}=\frac{c_{\rm p}T}{q}$, where $c_{\rm p}$ is the specific heat at constant pressure, $T$ is the temperature, and $q$ is the rate of energy generation due to the carbon burning) becomes comparable to the convective turnover timescale ($t_{\rm c}=\frac{H_{\rm p}}{u_{\rm c}}$, where $H_{\rm p}$ is the pressure scale height, and $u_{\rm c}$ is the convective velocity), e.g. $t_{\rm b}\sim{t_{\rm c}}$ (see also Woosley, Wunsch \& Kuhlen 2004; Lesaffre et al. 2006). During the late stage of carbon burning, the evolution of central core has experienced a phase that temperature increases sharply but density does not change any more. We adopt the position where ${T}_{\rm c}$ increases dramatically as the critical criterion of central explosive carbon ignition (i.e. green point in panel a of Fig.\,2, where $t_{\rm b}=\frac{1}{22}t_{\rm c}$); resulting in the formation of SNe Ia (the critical criterion for producing SNe Ia, see also Chen, Han \& Meng 2014; Wang, Podsiadlowski \& Han 2017; Wu et al. 2016, 2017).

In this simulation, we continue to evolve the CO core after ${T}_{\rm c}$ passes through the critical point until it increases to around red point, where $t_{\rm b}=\frac{1}{45}t_{\rm c}$. The accreting process has lasted for about $2\times10^{5}\,\mbox{yr}$, and finally the WD mass, central temperature and density are $1.3747{M}_\odot$, ${\rm log}T_{\rm c}\sim9.094$ and ${\rm log}{\rho}_{\rm c}\sim9.187$, respectively. The nuclear reaction rate in the centre at we stop our calculation is $\epsilon_{\rm nuc}=2.95\times10^{26}\,\mbox{erg}\,\mbox{g}^{-1}\,\mbox{s}^{-1}$, and over $86\%$ of the core in mass is in the convection zone.

\subsection{Oxygen-silicon cores}

\begin{figure}
\begin{center}
\epsfig{file=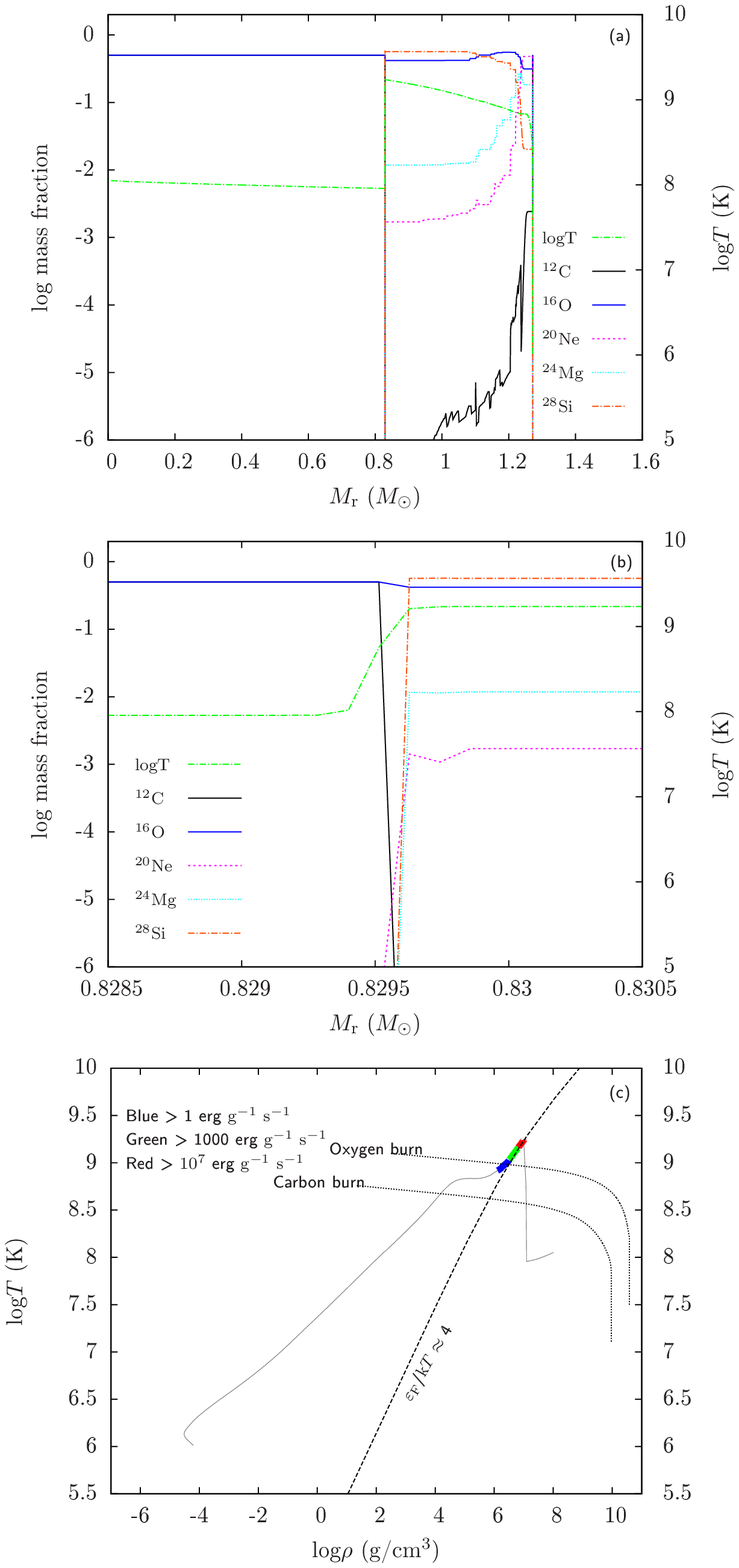,angle=0,width=11.9cm}
 \caption{An example of producing OSi cores, in which ${M}_{\rm WD}^{\rm i}=0.9{M}_\odot$ and $\dot{M}_{\rm acc}=4.0\times10^{-6}\,{M}_\odot\,\mbox{yr}^{-1}$. The carbon flame is propagating inwardly at this moment, and the burning ashes of carbon consist mainly of silicon. Panel (a): the elemental abundance distribution. Panel (b): the elemental abundance distribution near the flame. Panel (c): the density-temperature profile.}
  \end{center}
\end{figure}

If $\dot{M}_{\rm acc}$ is in the range of $\sim2.45-4.5\times10^{-6}\,{M}_\odot\,\mbox{yr}^{-1}$, the inwardly propagating carbon flames may lead to the formation of OSi cores directly. Fig.\,3 shows an example of producing OSi cores, in which ${M}_{\rm WD}^{\rm i}=0.9{M}_\odot$ and $\dot{M}_{\rm acc}=4.0\times10^{-6}\,{M}_\odot\,\mbox{yr}^{-1}$ (in the region of ``OSi cores''). In this case, carbon in the shell is ignited when the WD mass reaches $1.27{M}_\odot$. At the moment shown in Fig.\,3, the burning front has just passed the position where $M_{\rm r}=0.83{M}_\odot$. Through the abundance distribution profiles, we can see that the inwardly propagating burning wave transforms $^{\rm 12}{\rm C}$ into $^{\rm 20}{\rm Ne}$, $^{\rm 24}{\rm Mg}$ and $^{\rm 28}{\rm Si}$, through the nuclear reaction chains (1)-(3). However, the width of the flame is extremely thin ($\sim9.5\times10^{4}\,\mbox{cm}$) due to the high temperature of the burning front (${\rm log}T>9.2$; see panel (b) of Fig.\,3), resulting in that $^{\rm 20}{\rm Ne}$ is exhausted immediately once it appears through the nuclear reaction chains (4)-(6). As a result, the burning wave transforms $^{\rm 12}{\rm C}$ into $^{\rm 28}{\rm Si}$ directly in this simulation. Note that in this simulation we used a relative sparse computational grid to speed up the calculation, and in this case the flame interface may not be fully resolved (for more discussion see Sect. 4).

\begin{figure}
\begin{center}
\epsfig{file=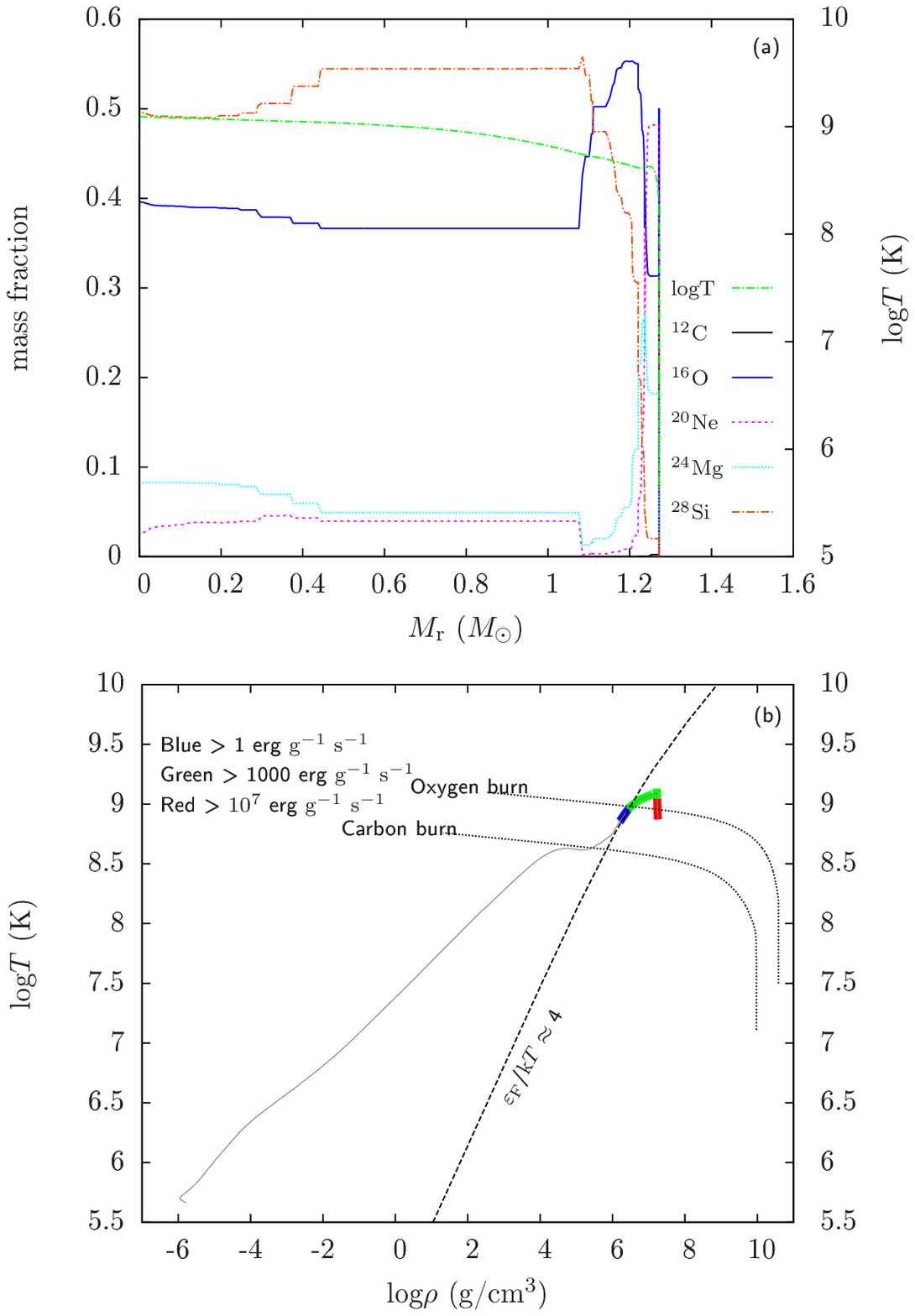,angle=0,width=12.2cm}
 \caption{Similar to Fig.\,3, but for the moment when the burning wave has reached the centre.}
  \end{center}
\end{figure}

The burning wave propagates into the core, and reaches the centre after more than $4000\,\mbox{yr}$. Fig.\,4 shows the elemental abundance distribution and $\rho$-$T$ diagram at this moment. The WD grows in mass up to $1.288{M}_\odot$ due to the mass-accretion process, and the mass fraction of $^{\rm 16}{\rm O}$ and $^{\rm 28}{\rm Si}$ in the central core are $40\%$ and $50\%$, respectively. The final outcomes of this OSi core can be divided into two cases as follows: (1) if the mass-accretion process is terminated, the core will cool down and become an OSi WD since silicon cannot be ignited under this $\rho$ and $T$; (2) if the accretion process continues, silicon flashes may occur when the core increases its mass up to $1.41{M}_\odot$, resulting in the formation of iron core, and a neutron star may be produced through the iron-core-collapse SN (e.g. Woosley \& Heger 2015; Schwab, Quataert \& Kasen 2016).

\subsection{Oxygen-neon cores}

\begin{figure}
\begin{center}
\epsfig{file=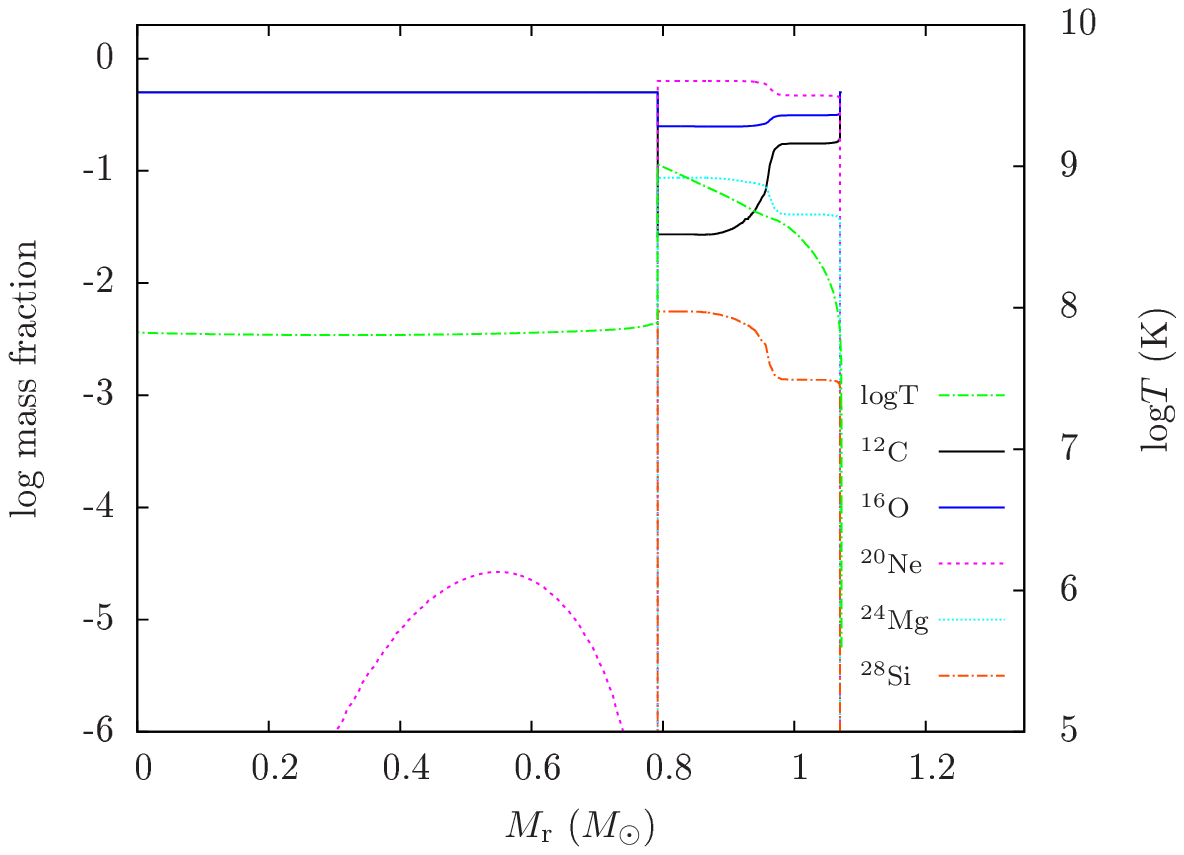,angle=0,width=12.2cm}
 \caption{The elemental abundance distribution of a $0.9{M}_\odot$ accreting CO WD during the process of surface carbon burning, in which $\dot{M}_{\rm acc}=1.0\times10^{-5}\,{M}_\odot\,\mbox{yr}^{-1}$. The inwardly propagating flame is transforming carbon into neon, and the CO core will evolve to an ONe core eventually.}
  \end{center}
\end{figure}

If $\dot{M}_{\rm acc}$ is in the range of $\sim4.5-1.05\times10^{-6}\,{M}_\odot\,\mbox{yr}^{-1}$, the inwardly propagating carbon flames may lead to the formation of ONe cores. Fig.\,5 presents the elemental abundance distribution of a $0.9{M}_\odot$ accreting CO WD during the surface carbon burning, in which $\dot{M}_{\rm acc}$ in this simulation is $1.0\times10^{-5}\,{M}_\odot\,\mbox{yr}^{-1}$ (in the region of ``ONe cores''). Carbon in the shell is ignited when the WD grows in mass up to $1.07{M}_\odot$ because of the high gravitational energy release rate. The temperature of the burning front is about $10^{9}{\rm K}$ which is lower than that in the example of OSi core. In this case, carbon in the shell will be transformed into $^{\rm 20}{\rm Ne}$ and $^{\rm 24}{\rm Mg}$ through the nuclear reactions chains (1) and (2). However, the temperature is not high enough for neon burning. The burning wave propagates inwardly and reach the centre after $2.4\times10^{4}\,\mbox{yr}$, leading to the formation of the ONe core. At this moment, the ONe core is surrounded by a thick CO envelope due to the accretion process, and the total mass of core $+$ envelope is $1.31{M}_\odot$.

\begin{figure}
\begin{center}
\epsfig{file=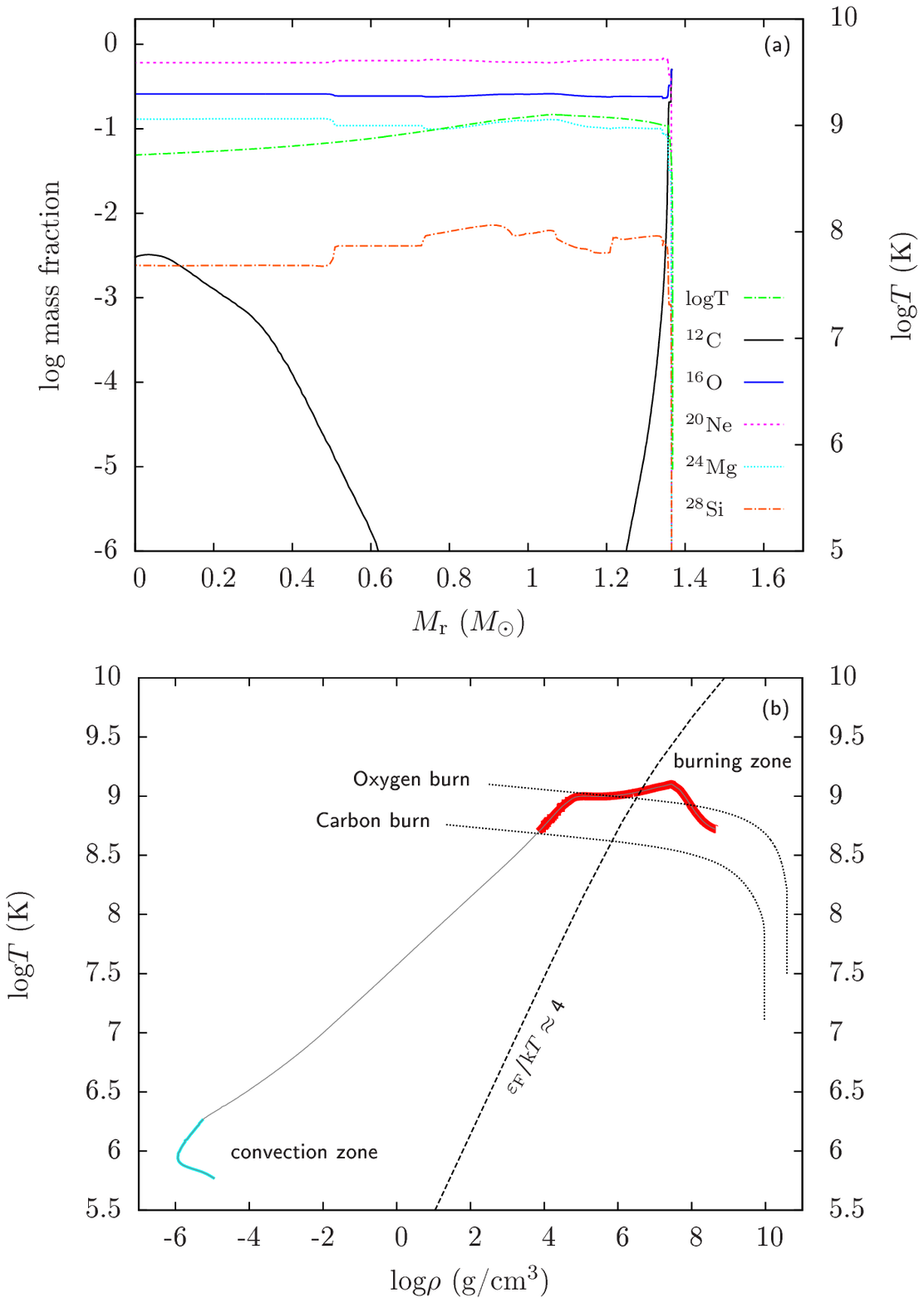,angle=0,width=12.2cm}
 \caption{An example of producing ONe cores, in which ${M}_{\rm WD}^{\rm i}=0.9M_\odot$ and $\dot{M}_{\rm acc}=1.0\times10^{-5}\,{M}_\odot\,\mbox{yr}^{-1}$. Panel (a): the elemental abundance distribution at the moment of carbon flame reaches the centre. Panel (b): the density-temperature profile at the moment of carbon flame reaches the centre.}
  \end{center}
\end{figure}

After the inwardly burning wave reaches the centre, the thick CO envelope will be ignited through an outwardly propagating burning wave. This process will last for $5000\,\mbox{yr}$, and finally a $1.368{M}_\odot$ ONe core is formed. Fig.\,6 presents the elemental abundance distribution and the $\rho$-$T$ diagram of the formed ONe core. The temperature of the core cannot trigger the off-centre neon burning in this case. Thus, the subsequent evolution of the ONe core can be divided into two cases as follows: (1) if the mass accretion-process is terminated and if the core can cool down, an ONe WD will form because the e-capture reactions of $^{\rm 24}{\rm Mg}$ and $^{\rm 20}{\rm Ne}$ cannot be triggered for such less massive core (the e-capture reaction of $^{\rm 24}{\rm Mg}$ occurs when ${\rm log}{\rho}_{\rm c}\sim9.6$, and the corresponding mass of ONe WD is about $1.381{M}_\odot$; e.g. Schwab, Quataert \& Bildsten 2015; Wu \& Wang 2018); (2) if ONe WD can form and the ONe WD continues to increase its mass, the occurrence of e-capture reactions will decrease the electron degenerate pressure of the core, leading to the formation of a neutron star through e-capture SN.

\subsection{Off-centre oxygen and neon ignitions}

If $\dot{M}_{\rm acc}$ is higher than $\sim1.05\times10^{-5}\,{M}_\odot\,\mbox{yr}^{-1}$, off-centre oxygen and neon will be ignited after the carbon flame reached the centre. Here, we present an example of producing off-centre oxygen and neon ignition, in which ${M}_{\rm WD}^{\rm i}=0.9{M}_\odot$ and $\dot{M}_{\rm acc}=1.5\times10^{-5}\,{M}_\odot\,\mbox{yr}^{-1}$ (in the region of ``off-centre O/Ne ignitions'').

\begin{figure}
\begin{center}
\epsfig{file=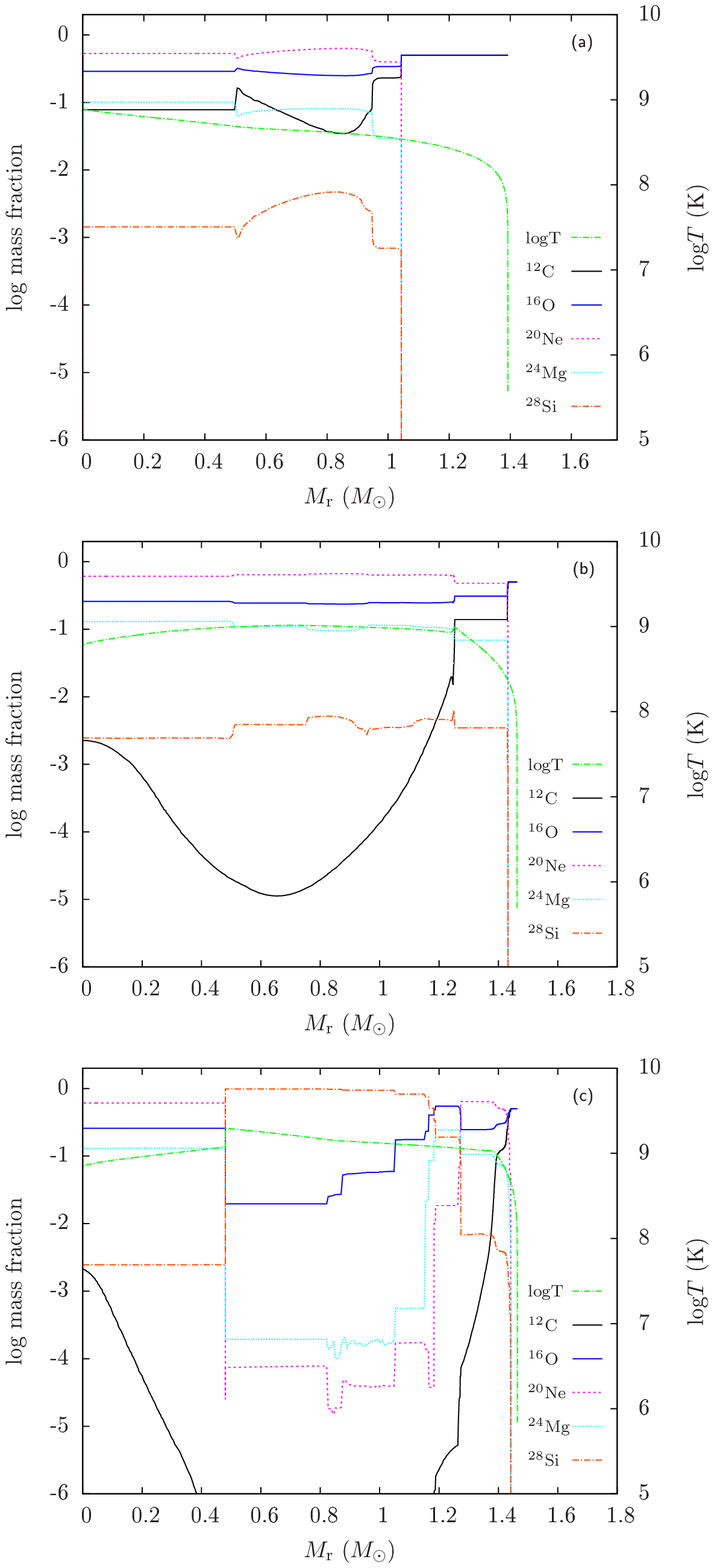,angle=0,width=11.9cm}
 \caption{An example of producing off-centre oxygen and neon ignition. Panels (a) - (c) present the elemental abundance distributions of a $0.9{M}_\odot$ accreting CO WD at different evolutionary stage, in which $\dot{M}_{\rm acc}=1.5\times10^{-5}\,{M}_\odot\,\mbox{yr}^{-1}$. Panel (a): when the carbon flame reaches the centre. Panel (b): when the outwardly propagating flame transforms all the carbon in the envelope into $^{\rm 20}{\rm Ne}$ and $^{\rm 24}{\rm Mg}$. Panel (c): during the stage of inwardly propagating neon flame.}
  \end{center}
\end{figure}

Fig.\,7 shows the elemental abundance distribution for the CO WD at different evolution stage. At the beginning of accreting process, CO material piles up onto the WD, leading to the formation of CO envelope. The off-centre carbon ignition occurs when the WD grows in mass up to $1.04{M}_\odot$, and after $23500\,\mbox{yr}$ the carbon flame reaches the centre. The total mass of the core $+$ envelope at this moment is $1.395{M}_\odot$ due to the continuous accretion process, in which the elemental abundance distribution is shown in panel (a). After the carbon flame reaches the centre, an outwardly propagating flame is triggered in the envelope. This phase has experienced $4533\,\mbox{yr}$, and a $1.46{M}_\odot$ ONe core is formed through the nuclear reaction chains (1)-(3), in which the elemental abundance distribution is shown in panel (b). The mass coordinate of the highest temperature point in the ONe core is $0.67{M}_\odot$, where the neon flame can be ignited in this position. The neon flame propagates inwardly, transforming $^{\rm 16}{\rm O}$, $^{\rm 20}{\rm Ne}$ and $^{\rm 24}{\rm Mg}$ into $^{\rm 28}{\rm Si}$ in a rapid way through the nuclear reaction chains (4)-(6), and the flame exhausts almost all the involved elements, leaving a Si envelope behind (mass fraction of $^{\rm 28}{\rm Si}\geq99\%$). The temperature of the burning front during the neon flame phase is ${\rm log}T\sim9.2-9.3$, in which the elemental abundance distribution is shown in panel (c).

\begin{figure}
\begin{center}
\epsfig{file=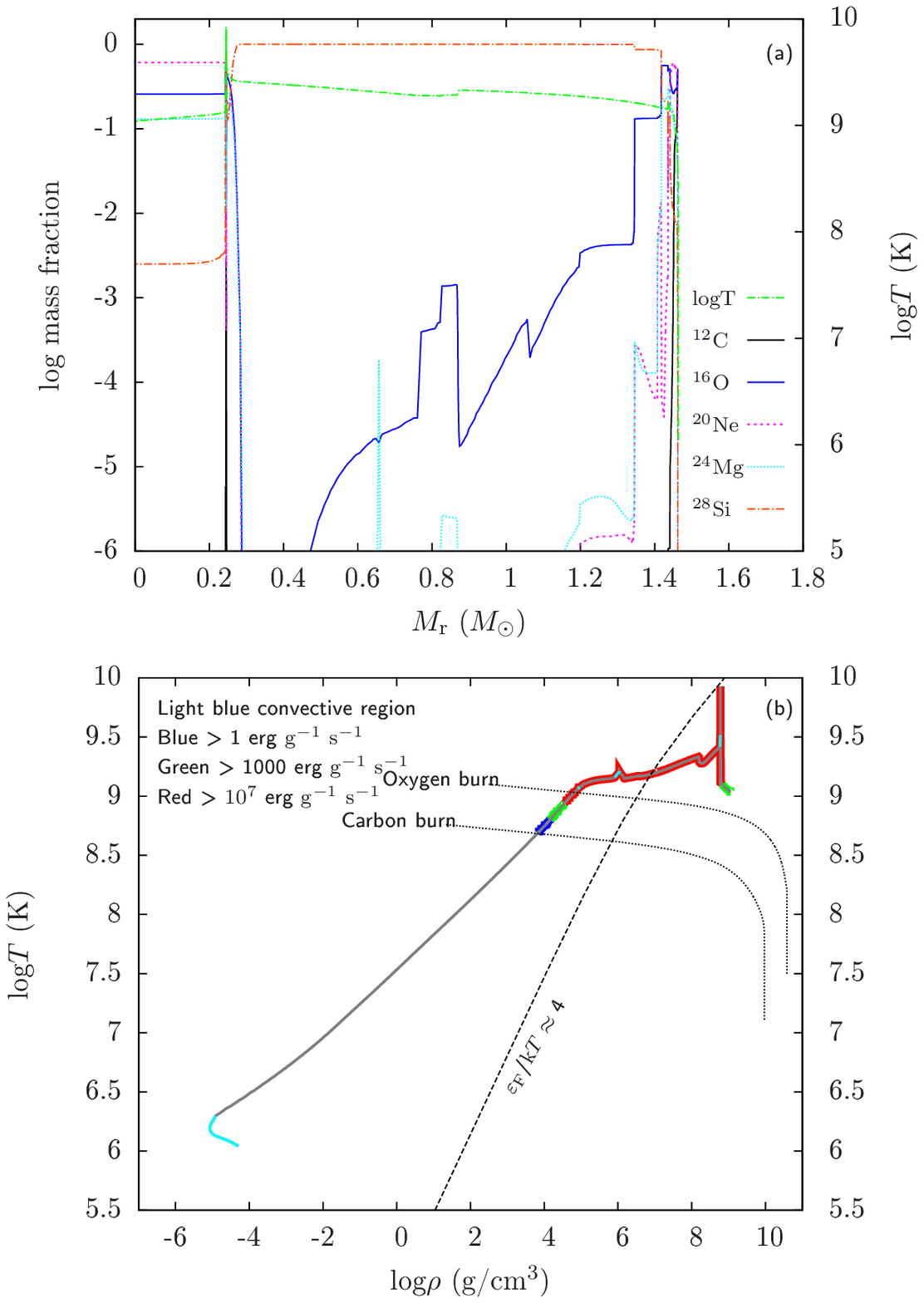,angle=0,width=12.2cm}
 \caption{An example of producing off-central dynamical process, in which ${M}_{\rm WD}^{\rm i}=0.9{M}_\odot$ and $\dot{M}_{\rm acc}=1.5\times10^{-5}\,{M}_\odot\,\mbox{yr}^{-1}$. Panel (a): the elemental abundance distribution at the moment of dynamical burning occurs. Panel (b): the density-temperature profile at the moment of dynamical burning occurs. The location of the flame during the explosion occurs is at ${M}_{\rm r}=0.245{M}_\odot$. The temperature and nuclear reaction rate at the position of the flame are ${\rm log}{T}_{\rm max}=9.93$ and $\epsilon_{\rm nuc}=4.95\times10^{26}\,\mbox{erg}\,\mbox{g}^{-1}\,\mbox{s}^{-1}$, respectively.}
  \end{center}
\end{figure}

Schwab, Quataert \& Kasen (2016) suggested that the inwardly propagating neon flame could reach the centre, resulting in the formation of Si core and subsequent Fe core or neutron star. However, in our simulations an dynamical process will occur when the neon flame reaches the position where mass coordinate is $0.245{M}_\odot$, in which the igniting situation could be slightly influenced by the choices of different nuclear networks (for more discussion see Sect. 4). The elemental abundance distribution and the $\rho-T$ profile of this simulation are shown in Fig.\,8. The temperature and nuclear reaction rate of the explosive ignition point at the moment of dynamical burning occurs are ${\rm log}{T}_{\rm max}=9.93$ and $\epsilon_{\rm nuc}=4.95\times10^{26}\,\mbox{erg}\,\mbox{g}^{-1}\,\mbox{s}^{-1}$, respectively. In this case, the high energy generation rate may trigger a deflagration or detonation wave, which propagates from the ignition point to the outside, burning the rest of the star. The gravitational binding energy may not bound such a high energy generation process. Thus, a huge matter ejection may be produced, resulting in a SN-like event. Marquardt et al. (2015) recently simulated the detonations of ONe WDs, and suggested that the synthetic light curves of ONe WD explosions can reproduce the luminous SN 1991T-like events.

\begin{figure}
\begin{center}
\epsfig{file=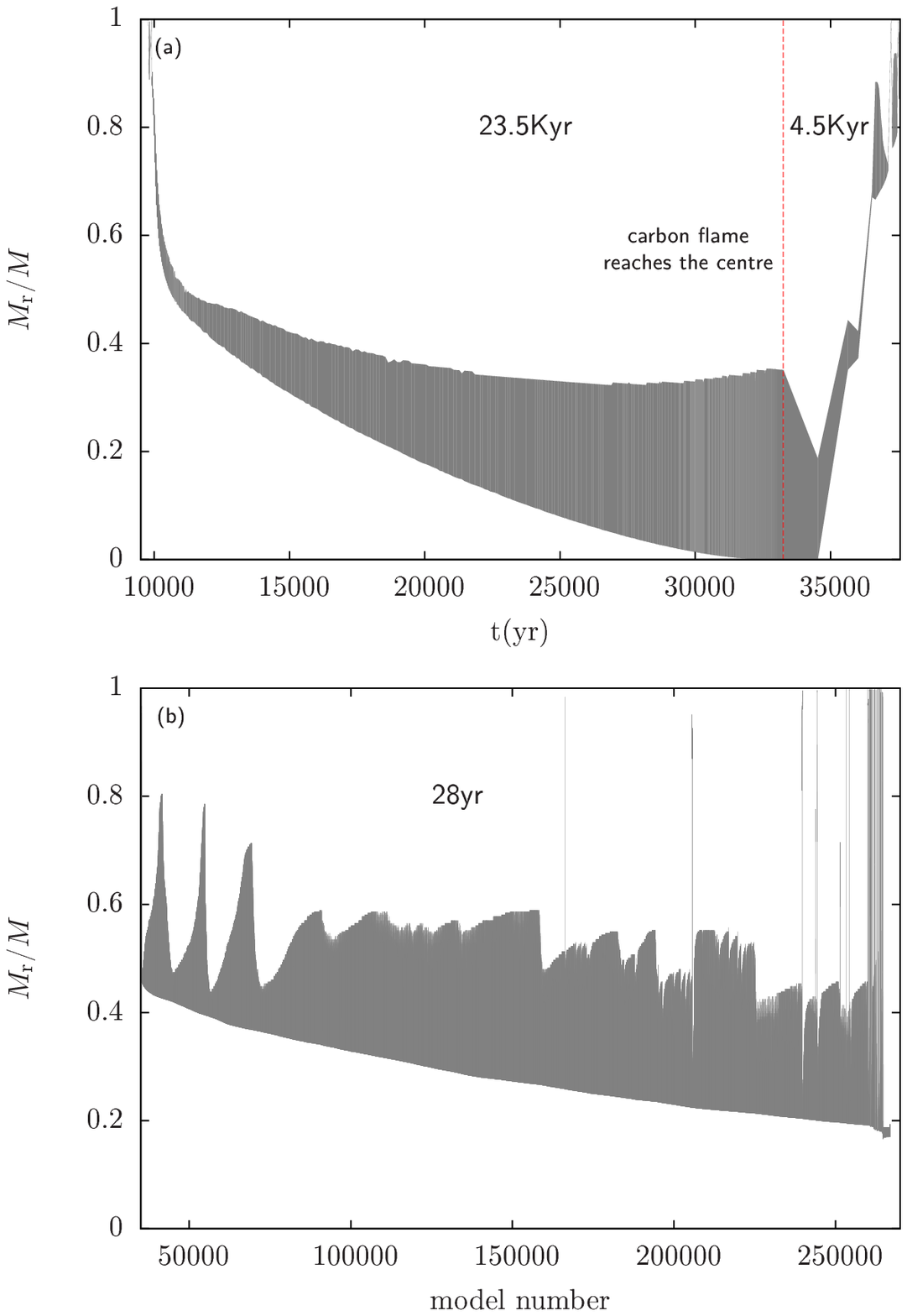,angle=0,width=12.2cm}
 \caption{Kippenhahn diagram of the $0.9{M}_\odot$ accreting CO WD during the evolution ($\dot{M}_{\rm acc}=1.5\times10^{-5}\,{M}_\odot\,\mbox{yr}^{-1}$). Panel (a): during the carbon burning phase, where the red dashed line presents the moment when the carbon flame reaches the centre. Panel (b): during the neon and oxygen burning phase. The evolution time for different phases are shown in both diagrams.}
  \end{center}
\end{figure}

In Fig.\,9, we present the Kippenhahn diagram during the evolution. The carbon flame experiences $2.35\times10^{4}\,\mbox{yr}$ to reach the centre, and after another $4.5\times10^{3}\,\mbox{yr}$, the CO WD is entirely transformed into ONe core. The neon flame propagates far more quickly then the carbon flame due to the high reaction rate, which experiences $28\,\mbox{yr}$ to trigger the dynamical process. From this figure, we can see that the carbon flame is relatively steady, whereas the neon flame moves inwardly in a fluctuant way. This can explain fluctuations of the elemental abundance distributions of $^{\rm 16}{\rm O}$, $^{\rm 20}{\rm Ne}$ and $^{\rm 24}{\rm Mg}$ in panel (a) of Fig.\,8.

\begin{figure}
\begin{center}
\epsfig{file=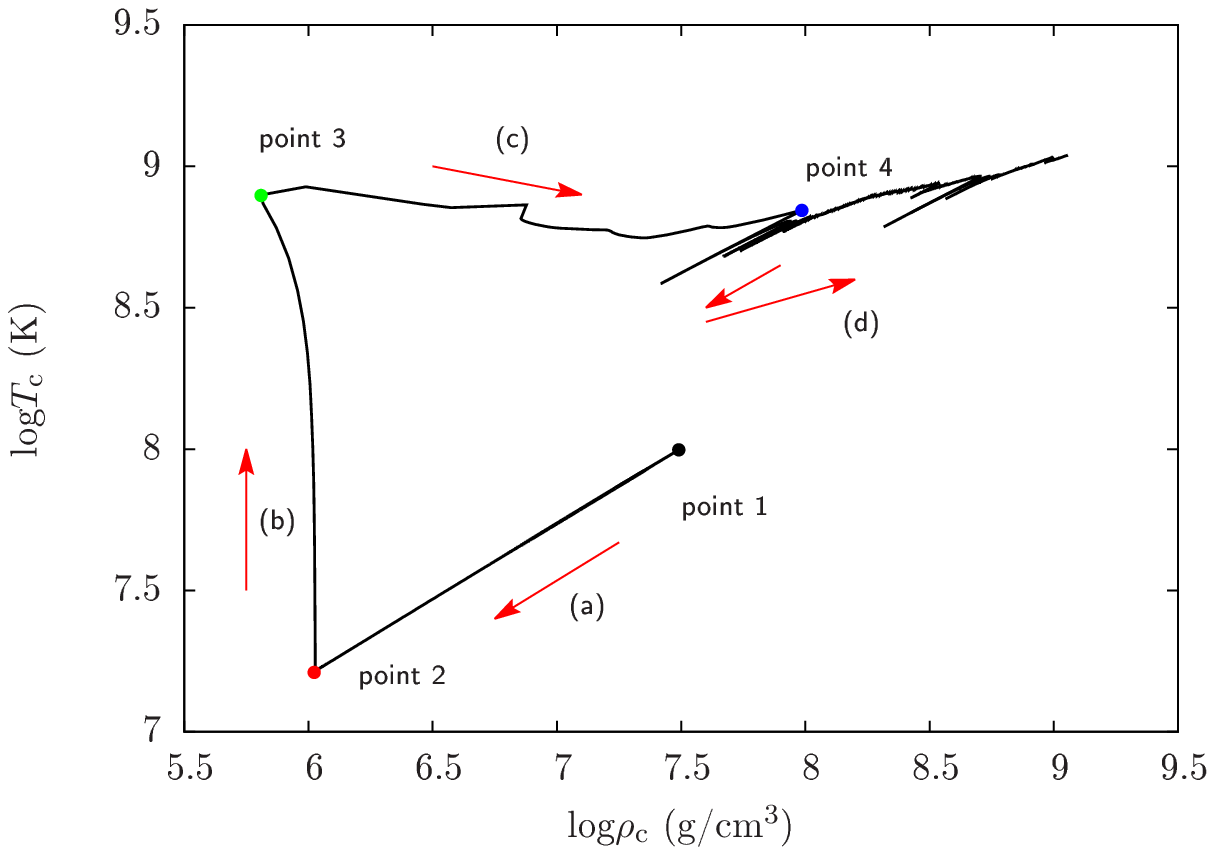,angle=0,width=12.2cm}
 \caption{Central density-temperature profile of the $0.9{M}_\odot$ accreting CO WD during the evolution, in which $\dot{M}_{\rm acc}=1.5\times10^{-5}\,{M}_\odot\,\mbox{yr}^{-1}$. Four points present different evolution phases. Point 1: carbon is ignited in the shell; point 2: the inwardly propagating carbon flame reaches the centre; point 3: the outwardly moving carbon flame reaches the surface; point 4: the off-centre neon ignition occurs. The red arrows point out the evolutionary direction of ${\rho}_{\rm c}$ and $T_{\rm c}$.}
  \end{center}
\end{figure}

The evolutionary track of ${\rho}_{\rm c}$ and $T_{\rm c}$ are shown in Fig.\,10. The evolutionary track can be divided into four parts based on the differently physical processes: (a) ${\rho}_{\rm c}$ and $T_{\rm c}$ rise to point 1 due to the mass-accretion process, where carbon in the shell is ignited. As the burning wave moves inwardly, the gravity of envelope applied to the core is decreased because of the radiation pressure produced by burning, resulting in ${\rho}_{\rm c}$ and $T_{\rm c}$ move to point 2; (b) when the burning wave reaches the centre, $T_{\rm c}$ increases dramatically due to the thermal energy release until it reaches point 3; (c) the carbon flame in the centre begins to quench gradually when the carbon in the thick envelope is ignited, leading to the increase of ${\rho}_{\rm c}$; (d) the surface neon is ignited at point 4. At this moment, the evolutionary track is similar to that of carbon burning. However, the propagation of neon flame is instable, resulting in the fluctuations of ${\rho}_{\rm c}$ and $T_{\rm c}$ in the evolutionary track.

\section{Discussion}\label{Discussion}

Nomoto \& Iben (1985) simulated the evolution of double CO WD mergers based on the thick-disc assumption, and found that the critical accretion rate of the off-centre carbon ignition $\dot{M}_{\rm cr}$ is in the range of $2-4\times10^{-6}\,{M}_\odot\,\mbox{yr}^{-1}$. Kawai \& Saio (1987) estimated the critical value is about $2.7\times10^{-6}\,{M}_\odot\,\mbox{yr}^{-1}$, which is irrespective of initial WD masses. Saio \& Nomoto (1985) suggested that CO WDs during merging may end up with single neutron stars if they experience off-centre carbon ignitions. In our simulations, off-centre carbon ignition can occur if $\dot{M}_{\rm acc}$ is higher than about $2.4\times10^{-6}\,{M}_\odot\,\mbox{yr}^{-1}$, which is a little bit lower than that of Kawai \& Saio (1987). However, we found that electron-capture induced collapse is not the only fate if $\dot{M}_{\rm acc}$ is higher than $\dot{M}_{\rm cr}$. Three conditions may occur (see Fig.\,1), which is different with previous works. Furthermore, off-centre carbon ignition could also occur in the CO WD+He star systems. Wang, Podsiadlowski \& Han (2017) simulated the long-term evolution of CO WDs accreting He-rich material, and found that off-centre carbon ignitions are triggered by surface steady helium burning. The critical value of He-accreting WDs is about $2.05\times10^{-6}\,{M}_\odot\,\mbox{yr}^{-1}$, which is lower than that in our simulations. This is because surface carbon is heated up by steady helium burning, resulting in that less gravitational force is needed to trigger carbon burning.

\begin{figure}
\begin{center}
\epsfig{file=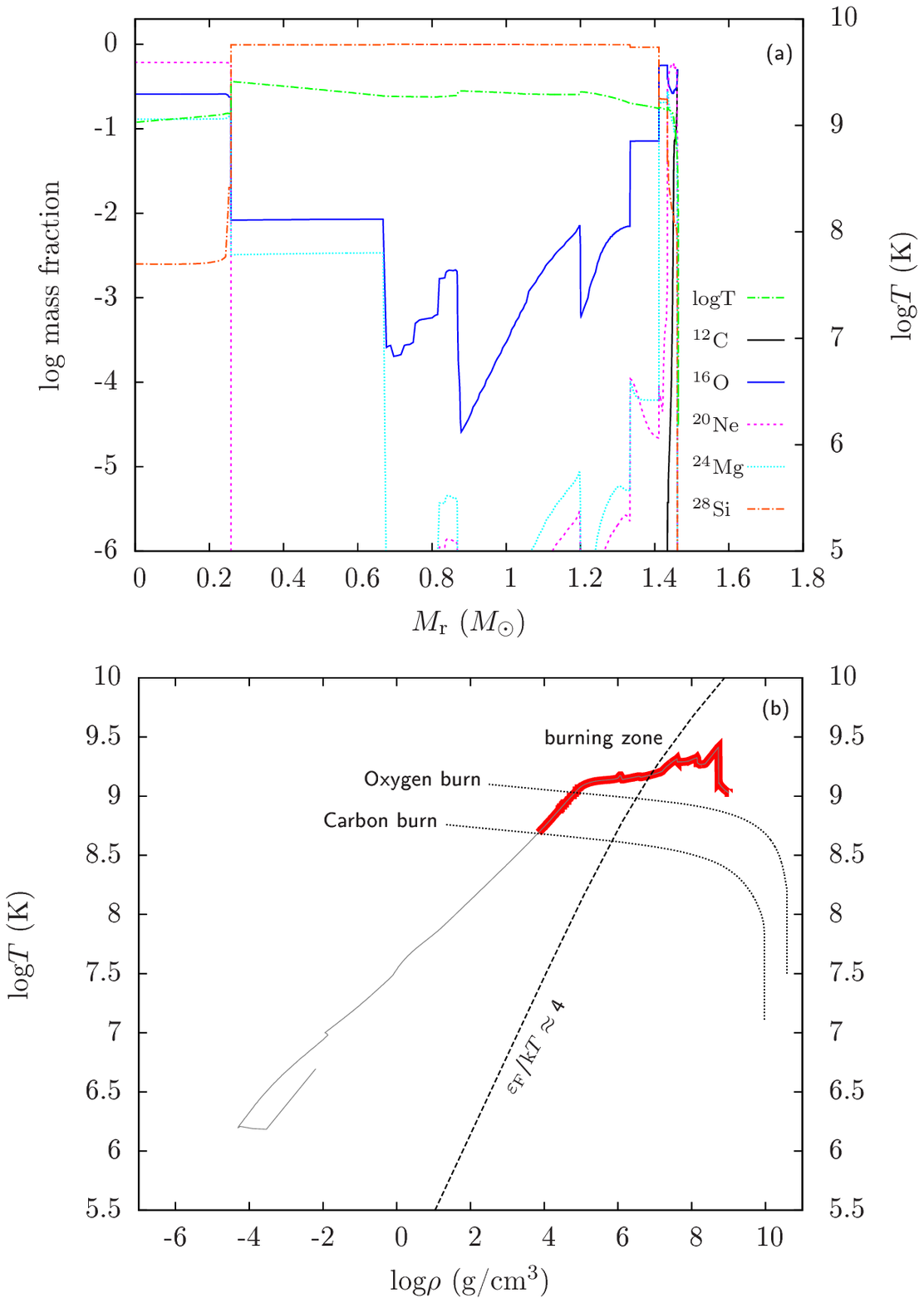,angle=0,width=12.2cm}
 \caption{Similar to Fig.\,8, but considering e-capture during the evolution.}
  \end{center}
\end{figure}

According to our simulations, the final outcomes of CO WDs may be neutron stars via e-capture reactions if $\dot{M}_{\rm acc}$ is in the range of ``ONe cores'' region (see Fig.\,1). However, in the region of ``off-centre O/Ne ignitions'', the core will also experience the stage of ONe core. Thus, it is necessary to investigate whether the e-capture reactions can influence the off-centre neon burning. In Fig.\,11, we present the elemental abundance distribution and the $\rho$-$T$ diagram for the accreting WD at the final stage of evolution by considering the e-capture reactions (i.e. $^{\rm 24}{\rm Mg}\,\rightarrow\,^{\rm 24}{\rm Na}\,\rightarrow\,^{\rm 24}{\rm Ne}$; $^{\rm 20}{\rm Ne}\,\rightarrow\,^{\rm 20}{\rm F}\,\rightarrow\,^{\rm 20}{\rm O}$). The evolutionary track of $T_{\rm c}$ and ${\rho}_{\rm c}$ in this simulation is similar to the ones without considering e-capture reactions. At the moment of off-centre neon ignition occurrence, ${\rm log}{\rho}_{\rm c}$ is approximately equal to $8$, which is lower than the threshold value of e-capture reactions of $^{\rm 24}{\rm Mg}$ (${\rm log}{\rho}_{\rm c}\geq9.6$; see Schwab, Quataert \& Bildsten 2015; Brooks et al. 2017). This result indicates that the e-capture reactions have little influence on the final outcome.

\begin{figure}
\begin{center}
\epsfig{file=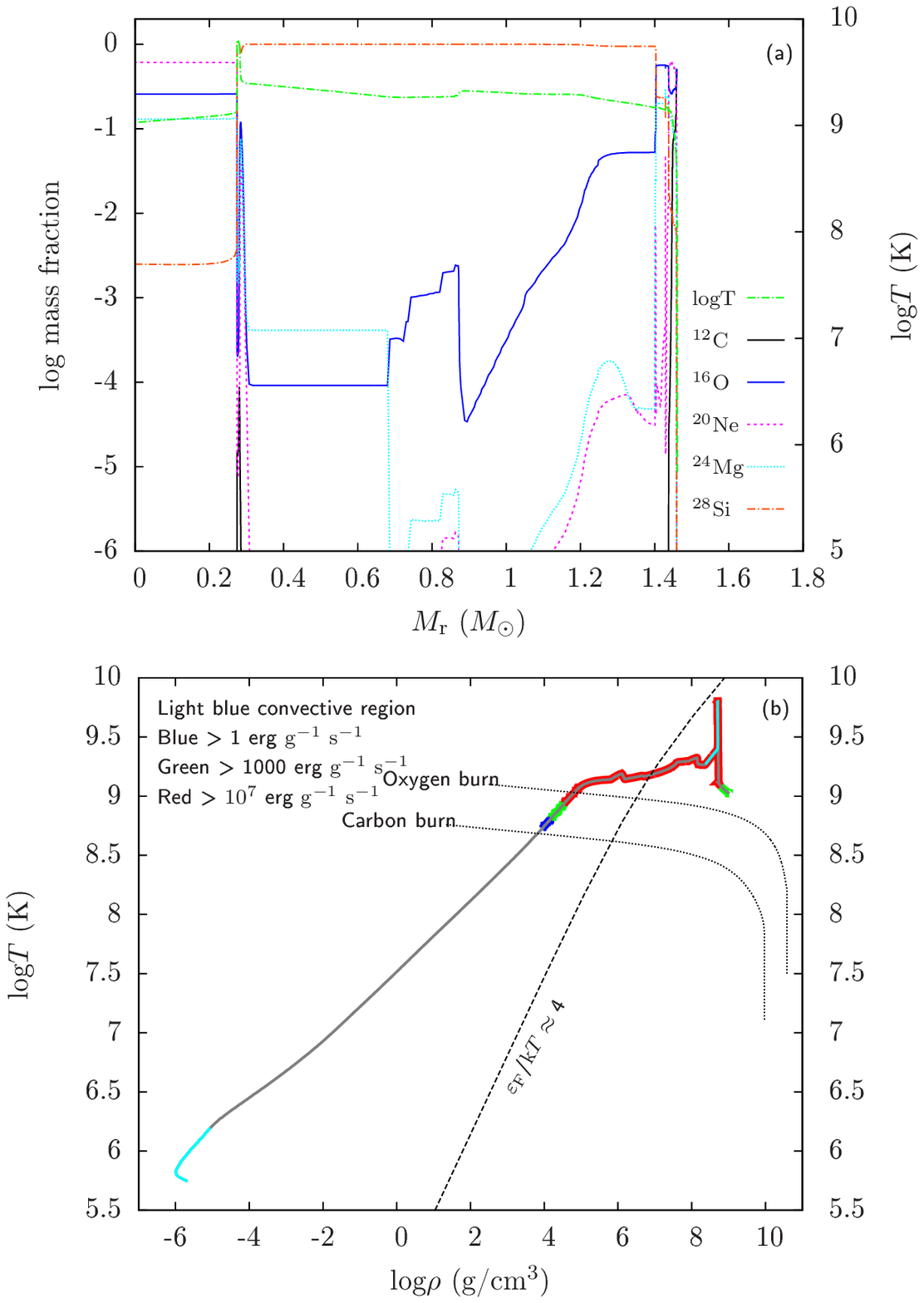,angle=0,width=12.2cm}
 \caption{Similar to Fig.\,8, but the cooling-time of initial CO WD is $t_{\rm cool}=3\times10^{8}\,\mbox{yr}$.}
  \end{center}
\end{figure}

\begin{table*}
\centering
\caption{Information for $0.9{M}_\odot$ accreting WD with different cooling time. Notes: ${M}_{\rm WD}^{\rm i}$ = initial WD mass; $t_{\rm cool}$ = cooling time; $T_{\rm eff}$ = effective temperature; $T_{\rm c}$ = central temperature; $\log\dot{M}_{\rm acc}$ = mass-accretion rate in logarithmic form; ${\rm P}_{\rm ig}$ = ignited position; ${M}_{\rm WD}^{\rm f}$ = the final mass of the core.}
\begin{tabular}{|c|c|c|c|c|c|c|}     % 8 columns
\hline
 ${M}_{\rm WD}^{\rm i} ({M}_\odot)$ & $t_{\rm cool} (\mbox{yr})$ & $T_{\rm eff} (\rm K)$ & $T_{\rm c} (\rm K)$ & $\log\dot{M}_{\rm acc} ({M}_\odot\,\mbox{yr}^{-1})$ & ${\rm P}_{\rm ig}$ & ${M}_{\rm WD}^{\rm f} ({M}_\odot)$\\
 \hline
 0.9 & $1.0\times10^{5}$ & $3.48\times10^{5}$ & $1.09\times10^{8}$ & $2.45\times10^{-6}$ & off-centre & $1.3781$\\
 0.9 & $1.0\times10^{5}$ & $3.48\times10^{5}$ & $1.09\times10^{8}$ & $2.40\times10^{-6}$ & centre &     $1.3745$\\
 0.9 & $1.0\times10^{6}$ & $2.33\times10^{5}$ & $7.34\times10^{7}$ & $2.45\times10^{-6}$ & off-centre & $1.3717$\\
 0.9 & $1.0\times10^{6}$ & $2.33\times10^{5}$ & $7.34\times10^{7}$ & $2.40\times10^{-6}$ & centre &     $1.3747$\\
 0.9 & $1.0\times10^{7}$ & $7.44\times10^{4}$ & $4.96\times10^{7}$ & $2.45\times10^{-6}$ & off-centre & $1.3717$\\
 0.9 & $1.0\times10^{7}$ & $7.44\times10^{4}$ & $4.96\times10^{7}$ & $2.40\times10^{-6}$ & centre &     $1.3748$\\
  \hline
\end{tabular}
\end{table*}

The CO WDs in our simulations have the same cooling time (i.e. $t_{\rm cool}=10^{6}\,\mbox{yr}$). However, the initial temperatures of accretion WDs may have influence on the mass-accretion processes and the subsequently evolutionary outcomes (e.g. Chen, Han \& Meng 2014). In order to investigate the effect of initial temperature, we simulated the long-term evolution of the accreting CO WDs with different cooling time and verified whether the initial temperature can influence the critical lines or not (for details, see Table\,2). In Fig.\,12, we present the elemental abundance distribution and $\rho$-$T$ diagram for the $0.9{M}_\odot$ accreting WD at the final evolutionary stage, in which $t_{\rm cool}=3\times10^{8}\,\mbox{yr}$. At the onset of accretion, $T_{\rm eff}$ and $T_{\rm c}$ are $1.47\times10^{4}\,{\rm K}$ and $1.43\times10^{7}\,{\rm K}$, respectively. This figure shows that the propagating process of flame and the final structure of the core are similar to the WD with $t_{\rm cool}=10^{6}\,\mbox{yr}$. The final core mass is $1.461{M}_\odot$ and the nuclear energy generation rate of the burning front is about $10^{22}\,\mbox{erg}\,\mbox{g}^{-1}\,\mbox{s}^{-1}$, which means that the dynamical process can still occur in this simulation. This result indicates that the cooling time has almost no influence on the final outcomes of accreting CO WDs.

\begin{figure}
\begin{center}
\epsfig{file=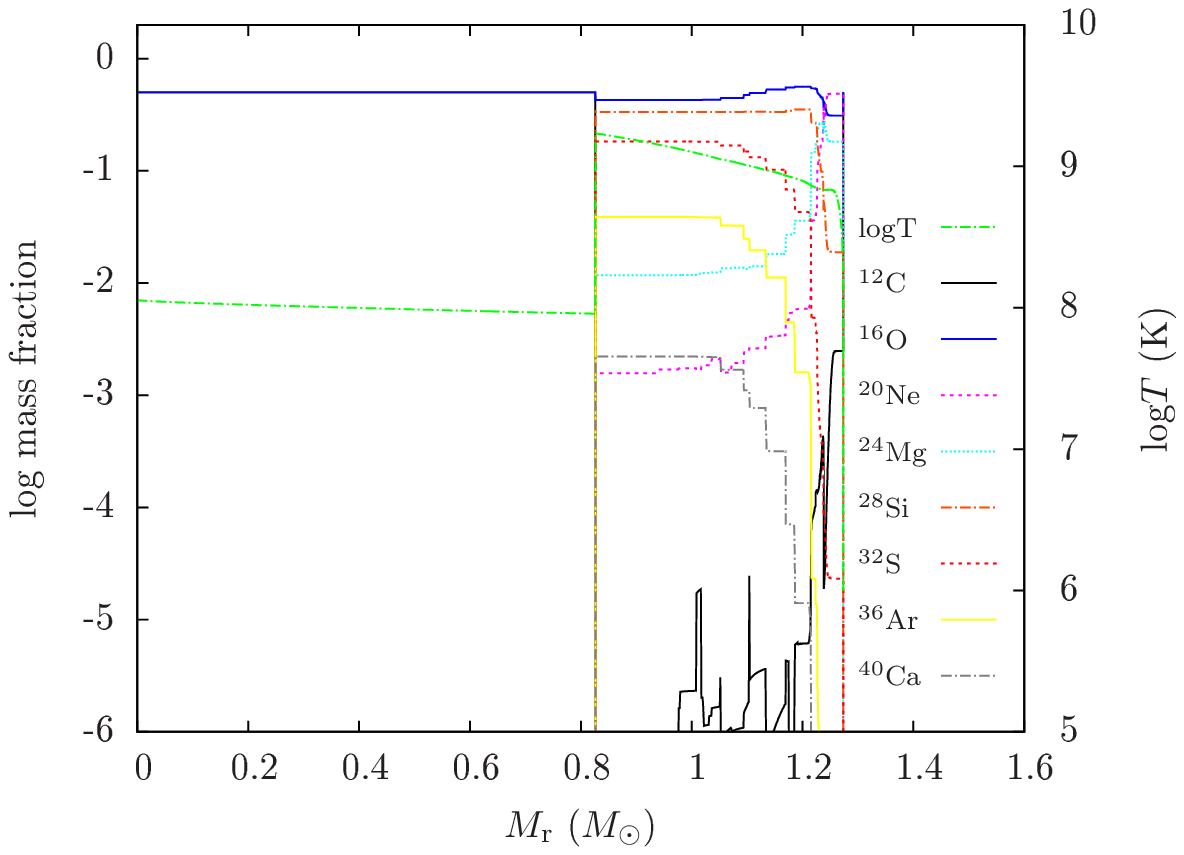,angle=0,width=12.2cm}
 \caption{Similar to panel (a) of Fig.\,3, but considering different nuclear reaction network.}
  \end{center}
\end{figure}

\begin{figure}
\begin{center}
\epsfig{file=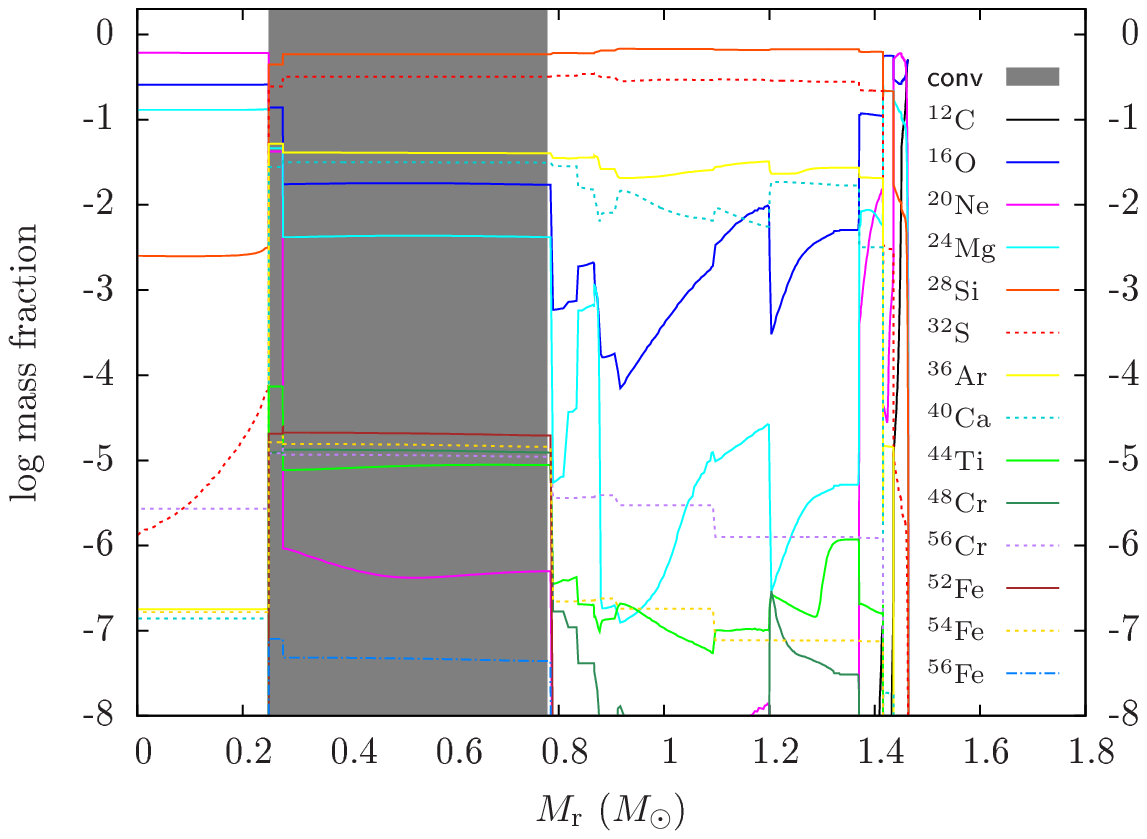,angle=0,width=12.2cm}
 \caption{Similar to panel (a) of Fig.\,8, but considering different nuclear reaction network.}
  \end{center}
\end{figure}

The nuclear reaction network used in our simulations is \texttt{co\_burn.net}, which does not include iron-group elements. To check whether different nuclear networks have influence on our results, we use \texttt{approx21.net} to recalculate our models. In this network, more isotopes are included, e.g. $^{\rm 32}{\rm S}$, $^{\rm 36}{\rm Ar}$, $^{\rm 40}{\rm Ca}$, $^{\rm 44}{\rm Ti}$, $^{\rm 48}{\rm Cr}$, $^{\rm 52-56}{\rm Fe}$, etc. In Fig.\,13, we present the elemental abundance distribution profile of a $0.9{M}_\odot$ accreting WD, in which we adopted the same initial parameters and assumptions as in Sect.\,3.2. The inwardly propagating flame in this simulation has reached the same position as in the panel (a) of Fig.\,3. For a comparison, the OSi envelope has almost the same O-Si ratio if we consider $^{\rm 28}{\rm Si}$ and $^{\rm 32}{\rm S}$ as the Si-group elements, which means that the influence of different nuclear networks on the result of ``OSi core'' is negligible. In Fig.\,14, we present the elemental abundance distribution profile of a $0.9{M}_\odot$ accreting WD, in which we adopted the same initial parameters and assumptions as in Sect.\,3.4. For a comparison, the evolution and structure of the core are similar before the dynamical process occurs. However, the dynamical burning condition is not as strong as that in Fig.\,8; this process produces a rapidly developing convective zone and releases energy outwardly, resulting in that the velocity of the surface becomes extremely high, which leads to a process of matter ejection. The final outcomes of the core after experiencing dynamical process may be divided into two possibilities if the flame can keep propagating inwardly: (1) if the dynamical process has not ejected too much mass, in this case the central density of the core has not decreased obviously, then an explosion may still occur at a position that is closer to the centre; (2) if the dynamical process leads to a relatively strong mass loss, and in this case the central density may decrease obviously, then the flame may propagate inside to the centre relatively stable, resulting in the formation of Si-Fe core surrounded by a thick Si-group elemental envelope. To understand whether the explosion could happen, dynamical simulations are needed in the future.

In order to accelerate the computation, we chose a relatively large timestep and sparse computational grid (i.e. ``${\rm varcontrol}\,\,{\rm target}=10^{-2}$'' and ``${\rm mesh}\,\,{\rm delta}\,\,{\rm coeff}=2.0$'') in our simulations. Note that the computation results may be influenced by the timestep and zoning (e.g. Woosley \& Heger 2015). We tested different choices of timestep and zoning for the examples of ``OSi cores'' and ``Off-centre O/Ne ignitions''. (1) For the example of ``OSi cores'' in Sect.\,3.2, we change the value of ``varcontrol target'' from ``$10^{-2}$'' to ``$10^{-4}$'' and ``mesh delta coeff'' from ``2.0'' to ``0.5''. In this case, the width and the propagating velocity of the flame are almost similar to that in Sect.\,3.2. We also used some additional controls to increase the number of grids at the flame interface (e.g. Farmer, Fields \& Timmes 2015). However, it is still difficult to fully resolve the flames in these MESA models. As indicated by the estimates in Woosley \& Heger (2015), resolving these flames can be quite challenging. (2) For the example of ``Off-centre O/Ne ignitions'' in Sect.\,3.4, we set the value of ``varcontrol target'' as ``$10^{-3}$'' and ``$10^{-4}$'', and ``mesh delta coeff'' as ``1.5'' and ``1.0''. We found that the dynamical process could also occur when the flame propagates to the similar position. These tests indicate that the different choices of timestep and zoning only have slightly influence on the final outcomes of CO cores in the present work.

After the carbon flame reaches the centre, the inwardly propagating neon flame would be triggered if the envelope mass is high enough. Schwab, Quataert \& Kasen (2016) investigated the post-merger evolution of two CO WDs, in which the masses of merger remnants are super $M_{\rm Ch}$ (i.e. $1.5$ and $1.6{M}_\odot$). They found that the off-central neon will be ignited if the mass of the formed ONe core is larger than $1.35{M}_\odot$, and the subsequent neon flame will reach the centre, resulting in the off-centre ignition of silicon if the mass of Si-rich core is larger than $1.41{M}_\odot$. In our simulations, the inwardly propagating neon flame would not reach the centre smoothly, i.e. a dynamical process will occur during the stage of neon burning, and the core mass when off-centre neon ignition occurs depends on the mass-accretion rates. These results imply that the final outcomes of double CO WD mergers are associated with the merging processes (e.g. slow merger, fast merger, composite merger, violent merger, etc.).

In the present work, we have not considered the influence of rotation on the long-term evolution of accreting CO WDs. Saio \& Nomoto (2004) investigated the evolution of CO WDs that accrete CO-rich material by taking into account the centrifugal force of rotation, and found that the cores are more massive when off-centre carbon ignition occurs, since the lifting effect produced by rotation reduces the density and temperature of the outer shells and delays the carbon ignition (see also Yoon \& Langer 2004). In our simulations, the velocity of burning front may be decreased and the carbon or neon ignited positions may be closer to the centre if considering the rotation. Meanwhile, the Urca-process cooling may lead to a lower $T_{\rm c}$ during the evolution of both CO and ONe core, especially when e-capture reactions of $^{\rm 24}{\rm Mg}$ occur (e.g. Schwab, Bildsten \& Quataert 2017). Jones et al. (2013) investigated the inwardly burning waves in super-AGB stars, and found that the neon burning wave may fail to reach the centre because of the Urca-process cooling, resulting in the formation of e-capture SNe rather than iron-core-collapse SNe. Schwab, Bildsten \& Quataert (2017) studied how this process affects the evolution of ${\rho}_{\rm c}$ and $T_{\rm c}$ for accreting ONe WDs, and suggested that the final outcomes of accreting ONe WDs would be accretion-induced collapse even though considering this cooling mechanism. This implies that the final outcomes of CO WDs in the region of ``off-centre O/Ne ignitions'' might be different, but for those in the region of ``O/Ne core'' might not be influenced if considering the Urca-process cooling.

\section{Summary}\label{Summary}

By employing the stellar evolution code MESA, we simulated the long-term evolution of CO WDs by accreting CO-rich material on the basis of the thick-disc assumption. We found that the off-centre carbon ignition will occur before the central carbon is ignited if the mass-accretion rate is higher than a critical value $\sim2.45\times10^{-6}\,{M}_\odot\,\mbox{yr}^{-1}$ for various WDs. Our works indicate that the final outcomes of WDs can be divided into four regions: SNe Ia, OSi cores then core-collapse SNe, ONe cores then e-capture SNe and off-centre O/Ne ignitions. We also found that the final outcomes are mainly determined by the mass accretion rates, whereas the influences of initial WD mass and cooling time can be ignored. In order to further understand the merging process of double CO WDs and subsequently evolution of the remnants, more CO WD pairs are hoped to be detected, and further numerical simulations are needed.

\section*{Acknowledgments}

We acknowledge the anonymous referee for the valuable comments. We thank Philipp Podsiadlowski, Zhanwen Han, Heran Xiong and Hailiang Chen for their helpful discussions.
This study is supported by the National Natural Science Foundation of China (Nos 11873085, 11673059, 11521303, 11390374 and 61561053),
the Chinese Academy of Sciences (Nos KJZD-EW-M06-01 and QYZDB-SSW-SYS001),
and the Natural Science Foundation of Yunnan Province (Nos 2018FB005 and 2017HC018).

%We acknowledge Stephen Justham, Xuefei Chen and Xiangcun Meng for their helpful discussions.
%We also thank Yan Gao for his kind help to improve the language of this paper.

\label{lastpage}
\end{document}